\documentclass[conference]{IEEEtran}
\IEEEoverridecommandlockouts
\usepackage{cite}
\usepackage{amsmath,amssymb,amsfonts}
\usepackage{algorithmic}
\usepackage{graphicx}
\usepackage{textcomp}
\usepackage{xcolor}
\usepackage[colorlinks=true, allcolors=blue]{hyperref}
\usepackage{subfig} 
 \usepackage{booktabs} 
 \usepackage{tabularx} 
 \usepackage{ragged2e} 
\usepackage{orcidlink} 
\usepackage{acronym} 
\usepackage[normalem]{ulem} 
\usepackage{xurl}   
\def\BibTeX{{\rm B\kern-.05em{\sc i\kern-.025em b}\kern-.08em
    T\kern-.1667em\lower.7ex\hbox{E}\kern-.125emX}}

\begin{document}

\title{Generative Explainability for Next-Generation Networks: LLM-Augmented XAI with Mutual Feature Interactions
\thanks{This work has been partially supported by the Swedish Innovation Agency (VINNOVA 2025-01348), and by the EUREKA CELTIC-NEXT project SUSTAINET-Advance funded by the Swiss Innovation Agency.}
\thanks{This is the authors' version of this paper. The final edited version is available at \url{https://doi.org/10.1109/WiMob66857.2025.11257542}.}
\thanks{The implementation of this work is available at \url{https://github.com/kiarashRezaei/llm-for-xai-qotEstimation}.}
}

\author{
    \IEEEauthorblockN{Kiarash Rezaei\IEEEauthorrefmark{1}\orcidlink{0009-0003-6166-2614}, Omran Ayoub\IEEEauthorrefmark{2}\orcidlink{0000-0002-3884-3594}, Sebastian Troia\IEEEauthorrefmark{3}\orcidlink{0000-0002-4712-3767}, Francesco Lelli\IEEEauthorrefmark{2}\IEEEauthorrefmark{4}\orcidlink{0000-0003-1900-9171},\\ Paolo Monti\IEEEauthorrefmark{1}\orcidlink{0000-0002-5636-9910}, Carlos Natalino\IEEEauthorrefmark{1}\orcidlink{0000-0001-7501-5547}}
   \\ \IEEEauthorblockA{\IEEEauthorrefmark{1} Department of Electrical Engineering, Chalmers University of Technology, 412 96 Gothenburg, Sweden
    \\ \{kiarashr, mpaolo, carlos.natalino\}@chalmers.se}
    \IEEEauthorblockA{\IEEEauthorrefmark{2}University of Applied Sciences and Arts of Southern Switzerland, 6928 Lugano, Switzerland
    \\ omran.ayoub@supsi.ch}
    \IEEEauthorblockA{\IEEEauthorrefmark{3}Politecnico di Milano, 20133 Milan, Italy
    \\ sebastian.troia@polimi.it}
    \IEEEauthorblockA{\IEEEauthorrefmark{4}Tilburg University, 5037 AB, Tilburg, Netherlands.
    \\f.lelli@tilburguniversity.edu}
    }

\maketitle

\begin{abstract}
As \ac{AI/ML} models become integral to network operations, their lack of transparency poses a significant barrier to operator trust. 
Existing \ac{XAI} techniques often fail to bridge this gap for non-specialists, producing technical outputs that are difficult to translate into actionable insights. 
This paper presents a framework specifically designed to address this shortcoming. 
It leverages a moderately sized \ac{LLM} and extends beyond the standard use of \ac{SHAP} feature influence values. 
The framework employs a structured prompt enriched with mutual feature interaction data to generate human-understandable natural language explanations. 
To validate our framework, we performed an empirical evaluation on an optical \ac{QoT} estimation use case with human evaluators. 
We collected independent performance evaluations from specialists, which showed a high inter-evaluator agreement. 
Compared to a state-of-the-art baseline that uses only SHAP feature influence values in a straightforward prompt, our approach improves the explanation usefulness and scope by 12.2\% and 6.2\%, while achieving 97.5\% correctness.
\end{abstract}


\section{Introduction}
Modern communication systems and \acf{AI} are increasingly intertwined. 
The rapid growth of data transport demands and widespread adoption of AI-driven applications have transformed communication networks into critical enablers of contemporary digital services \cite{Alhammadi2024}.
Managing such systems is complex, posing significant challenges in network automation tasks related to resource allocation, fault detection, and maintaining optimal performance under dynamic conditions \cite{nyswa2025future}. 
Hence, \ac{AI/ML} are increasingly employed to tackle these issues by providing data-driven solutions that aim at network efficiency and reliability. 
The broad applicability of these techniques is demonstrated by their use across diverse networking domains, from wireless mobility management and cellular resource allocation \cite{El-Hajj2025} to optical network resource allocation and management \cite{Musumeci_2019_survey,Natalino:24,Etezadi:23}.

As AI/ML models become more integrated into network automation workflows, the need for transparency and interpretability of them has become critical \cite{wu2022knowledge, wang2021applications, ayoub2022towards, ayoub2024xrl}. 
Explainable artificial intelligence (XAI) techniques, such as \ac{SHAP} \cite{lundberg2017unifiedapproachinterpretingmodel}, have been utilized to interpret model predictions and provide insights into model behavior.
\Ac{XAI} is essential before model deployment (to inspect valid model behavior) and in scenarios where human supervision and/or intervention may be required to validate or override automated decisions \cite{dutta2021challenge}. 
In the context of network automation, an engineer may need to inspect specific decisions made by \ac{AI/ML}-based systems, particularly when the system takes an action that appears counterintuitive, such as rerouting traffic through a seemingly congested path. 
In such cases, explainability mechanisms are critical for understanding the rationale behind the decision, verifying that it aligns with operational policies or inferred predictions, and deciding whether to trust the model or override its action manually. 
Nonetheless, the explanations extracted from \ac{XAI} techniques are often presented in technical formats that are cognitively demanding. 
They also require domain-specific expertise to be interpreted \cite{hudon2021explainable, weber2023beyond}, posing a challenge for their broader adoption. 
The literature highlights this difficulty~\cite{ahmed2024explainable, bikkasani2024ai}, showing that while standard \ac{XAI} methods can rank feature importance, they struggle to translate these rankings into contextually meaningful and actionable insights. 

\begin{figure}[t] 
    \centering
    \begin{minipage}[b]{0.49\linewidth}
        \centering
        \includegraphics[width=\linewidth]{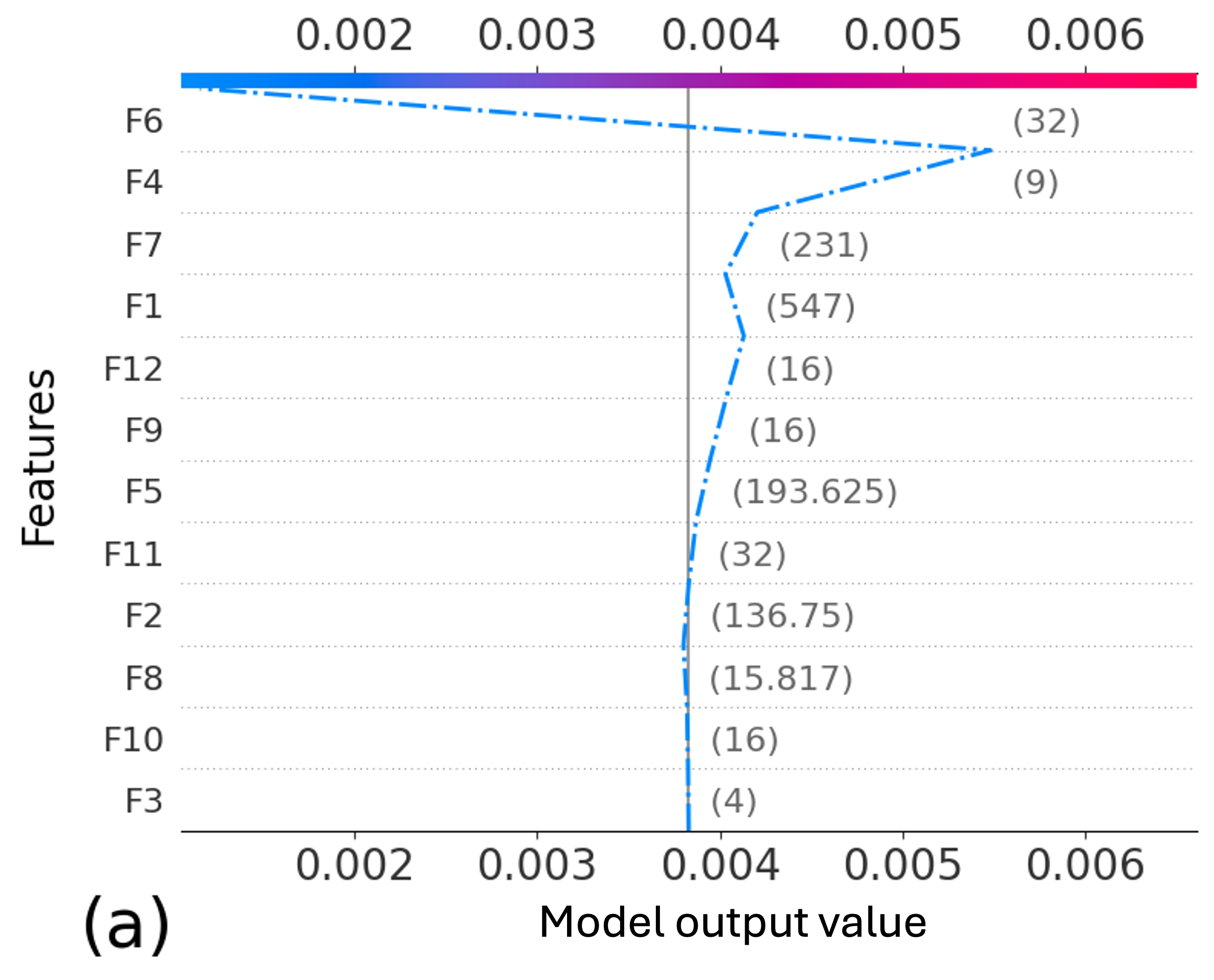}
        \label{fig:shap_influence}
    \end{minipage}
    \hfill
    \begin{minipage}[b]{0.49\linewidth}
        \centering
        \includegraphics[width=\linewidth]{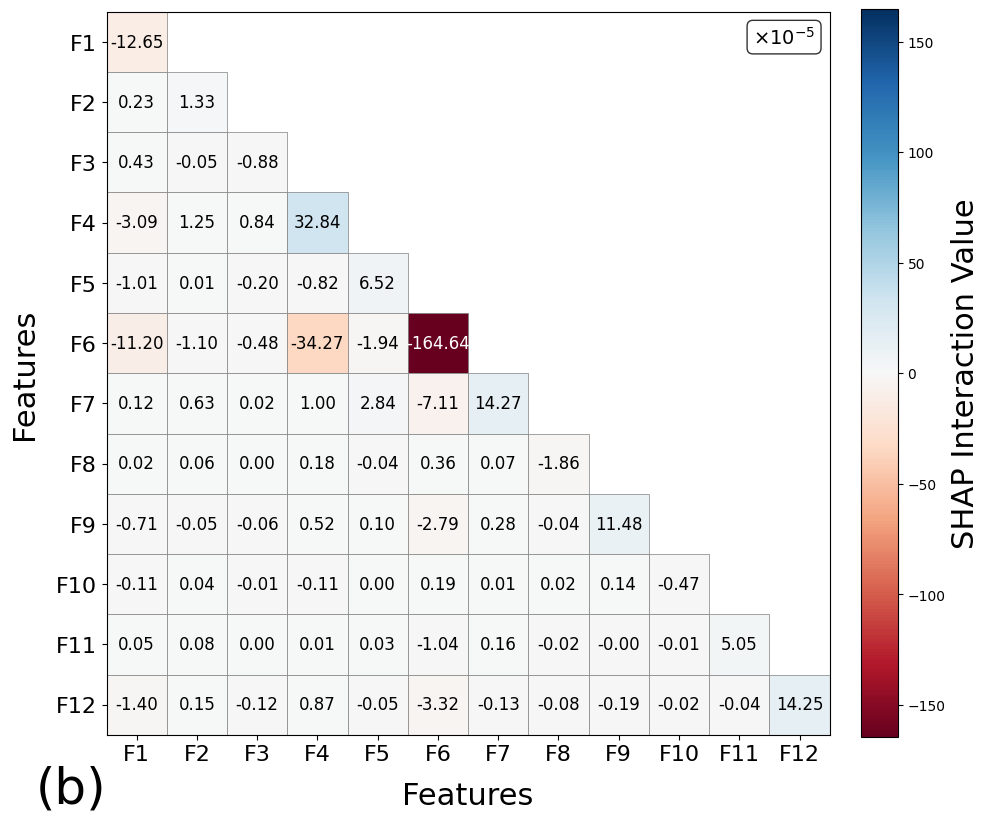}
        \label{fig:shap_interaction}
    \end{minipage}
    \caption{Illustrative example of SHAP-based explanation for a hypothetical QoT estimation instance with 12 features: 
    (a) Feature influence plot showing individual feature contributions. 
    (b) Feature interaction heatmap illustrating pairwise effects.}
    \label{fig:shap}
\end{figure}

Figure \ref{fig:shap}(a) shows an illustrative example of SHAP feature influence values, where $F_1$..$F_{12}$ represent arbitrary features. 
They quantify the isolated, individual contribution of each one of them to the model's output. 
Similarly, Fig. \ref{fig:shap}(b) shows an example of SHAP mutual feature interaction values. 
These values quantify the synergistic effect where the combined impact of two features differs from the sum of their individual influences. 
Both figures contain a high volume of raw, multidimensional data, including magnitude and directional influence, which makes it difficult to accurately distill the model's core logic without significant effort and/or domain knowledge (from both \ac{XAI} and the technical details of the specific use case).

In this work, we propose a framework designed to address the challenge of interpreting complex \ac{XAI} outputs.
Our framework leverages modern \acp{LLM} and the nuanced insights from mutual feature interaction values to generate human-understandable explanations. 
These explanations enhance the transparency (and perceived trustworthiness) of AI/ML models while significantly reducing the cognitive load, time, and expertise required by a human engineer to interpret \ac{XAI} outputs correctly.
As a case study, we focus on the \acf{QoT} estimation problem in optical networks \cite{rottondi2018machine, allogba2022machine}. 
This problem involves predicting whether a candidate lightpath will meet predefined signal quality thresholds (e.g., bit error rate) before it is provisioned (i.e., before it is allocated in the network). 
In practical scenarios, receiving clear explanations for \ac{QoT}-related decisions is highly valuable for network engineers, especially when the model deems a particular lightpath as unacceptable \cite{ayoub2022quantifying, fawaz}. 
For instance, if a proposed route is rejected despite appearing viable based on traditional engineering rules, the operator must understand the model's reasoning to assess whether the decision is based on subtle patterns (e.g., cumulative non-linear penalties or latent correlations with adjacent lightpaths) or is potentially the result of model overfitting or error. 
Human-readable explanations in such cases support trust, allow manual validation or override when needed, and ultimately lead to safer and more efficient network operation.

Empirical results validate the proposed framework. By guiding a reasoning \ac{LLM} with a structured prompt that incorporates both \ac{SHAP} feature influence and mutual feature interaction values, we produce higher-quality explanations. To assess these explanations, we adopt human-centered evaluation metrics that capture how correct, comprehensive, and useful the generated explanations are from the perspective of network engineers. Compared to a state-of-the-art approach using only plain \ac{SHAP} feature influences, our framework demonstrates significant improvements across these metrics: usefulness increased by 12.2\%, scope by 6.2\%, and correctness reached 97.5\%.
While the core of this work is an \ac{XAI}-enhanced framework, its motivation lies in network automation, where timely and reliable decisions are essential \cite{nyswa2025future, Musumeci_2019_survey}. Our case study addresses \ac{QoT} estimation, yet the framework applies broadly wherever \ac{AI/ML} models require transparency. This includes tasks such as traffic engineering, mobility management, slicing, and fault recovery \cite{Alhammadi2024, adanza2024intent}, where explainability helps engineers interpret AI decisions, validate them against policies, and preserve human oversight.

\section{Related Work}

Recent research addressed the gap between technical model outputs and user comprehension by proposing the integration of \acp{LLM} and \acs{AI/ML} to help translate complex model explanations into human-interpretable language, 
Zeng et al. \cite{zeng2024enhancinginterpretabilityshapvalues} demonstrated that \acp{LLM} can effectively translate SHAP value outputs into natural language explanations. 
This approach has shown particular promise across diverse application domains, with studies such as \cite{nazi2024largelanguagemodelshealthcare} examining how \acp{LLM} can generate near-human-level explanations for healthcare applications, enhancing both interpretability and clinical decision-making processes. 
Building on these results, research in communication networks has specifically investigated the integration of \acp{LLM} with \ac{SHAP} methodologies to enhance the interpretability of \acs{AI/ML} models deployed in network automation systems. 
For instance, a pipeline combining anomaly detection, SHAP-based root cause analysis, and \ac{LLM}–generated corrective actions was proposed for managing 6G microservices environments \cite{10742571}. 
However, this evaluation relied primarily on automatic metrics (e.g., BERT, ROUGE) that assess surface-level properties like text similarity. they achieved a BERT score around 0.74 out of 1.0 using Llama2 model with 70 billion parameters.
In the context of \ac{QoT} estimation, ChatGPT 3.5 with 175 billion parameters has been used to enhance SHAP explanations through direct prompting and self-reflection \cite{Ayoub_2025_icton}, although without a structured prompt design.
It achieved, with self-reflection, a correctness of 65\%.
Both works show a relatively low score for the explanations generated by the \acp{LLM}. 
Moreover, both works utilize traditional \acp{LLM} without reasoning capabilities and rely solely on isolated SHAP values as a measure of feature importance. 
No importance is given to mutual feature interactions that can potentially show how features jointly affect the ML model predictions.

Based on the above considerations, this paper considers two research questions: \emph{(i)} do reasoning \acp{LLM} improve the performance of explanations?; and \emph{(ii)} can the performance of \acp{LLM} in generating interpretations be enhanced by incorporating SHAP feature interaction values within a structured prompt?
To answer these questions, the paper presents a framework with two key advancements over the state-of-the-art. 
First, to explore the impact of \ac{LLM} capability, we harness the power of a reasoning \ac{LLM} with a moderate number of parameters. Second, to test the value of deeper contextual data, we enrich our structured prompt with \ac{SHAP} mutual feature interaction values. While these values are often disregarded due to their complex characteristics (as illustrated in Fig.~\ref{fig:shap}(b)), our assumption is that they provide critical context for the \ac{LLM}'s interpretation process, positioning our work beyond prior studies such as \cite{Ayoub_2025_icton} by combining reasoning \acp{LLM} with structured feature-interaction-aware context.

\section{LLM-Augmented XAI Framework with Feature Interactions}
\label{sec: llm-augmented xai framework}

\begin{figure*}[t]
   \centering
    \includegraphics[width=\textwidth]{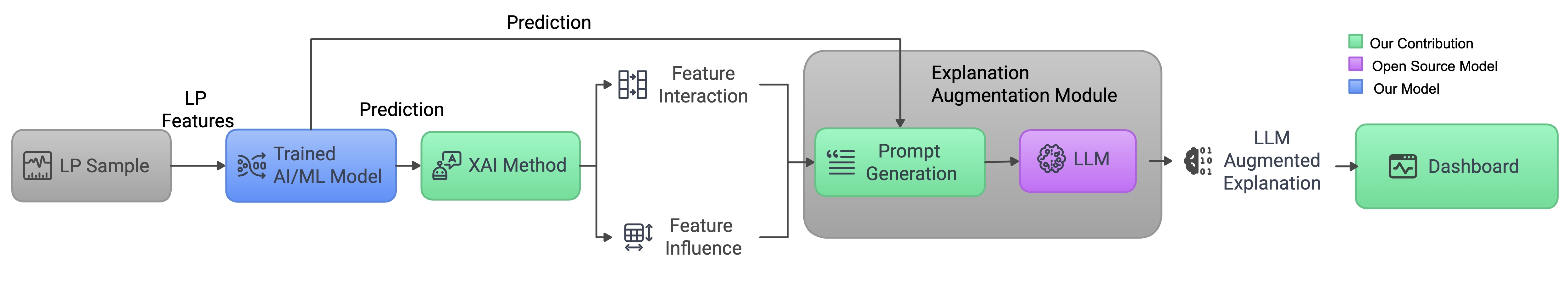}
    \caption{LLM-Augmented XAI framework pipeline: from input sample processing (left) to the generation of \ac{LLM} augmented explanations (right). Key components and contributions are color-coded.}
    \label{fig:framework}
\end{figure*}

The proposed framework comprises four parts: \emph{(1)} AI/ML model, \emph{(2)} \ac{XAI} method, \emph{(3)} explanation augmentation module, and \emph{(4)} dashboard (Fig.~\ref{fig:framework}). 
The goal is to enhance the interpretability of AI/ML model decisions through an LLM by processing the feature contribution values extracted via SHAP.

\noindent
\textit{\uline{AI/ML Model:}}
The first component is the AI/ML model to be explained. 
The use case assumed for this work is the \ac{QoT} estimation of optical lightpaths. 
It was selected primarily for its practical importance in ensuring service reliability and building operator trust. 
The technically challenging nature of the task, characterized by complex non-linear feature interactions, serves as an additional benefit, providing a robust testbed for our framework. 
To this end, we employ an \ac{XGB} model, given its proven effectiveness in prior works \cite{Ayoub_2025_icton}. 
The model takes as input a set of features describing a candidate lightpath and outputs the estimated \ac{BER}.
It is important to note that the proposed framework remains model-agnostic, i.e., it is applicable to various AI/ML models and tasks.

\noindent
\textit{\uline{XAI Method:}} 
To explain the model's decisions, we apply SHAP~\cite{lundberg2017unifiedapproachinterpretingmodel}, a method for post-hoc explainability quantifying local feature contributions using Shapley values.
This method allows for quantifying both the individual importance of features—capturing their mutual influence on the model prediction (Fig. \ref{fig:shap}(a))—and their mutual interactions, which reflect the joint impact of feature pairs (Fig. \ref{fig:shap}(b)). 
Together, these outputs serve as the core inputs for the explanation augmentation module. 
We focus on local explanations, which provide insight into an individual model's prediction. 
This is particularly relevant in an AI/ML-aided scenario where an expert reviews the model's outputs before they are adopted in production.

\noindent
\textit{\uline{Explanation augmentation module:}} The key element of this module is the structured prompt, detailed in Table~\ref{tab:prompt_structure} in the Appendix, which instructs the \ac{LLM} in generating augmented explanations. 
This prompt incorporates the model prediction and the individual feature influence values as primary content. 
It is then further enriched with mutual feature interaction scores to capture the more subtle aspects of the model's behavior.

\begin{figure}[t]
   \centering
    \includegraphics[width=\linewidth]{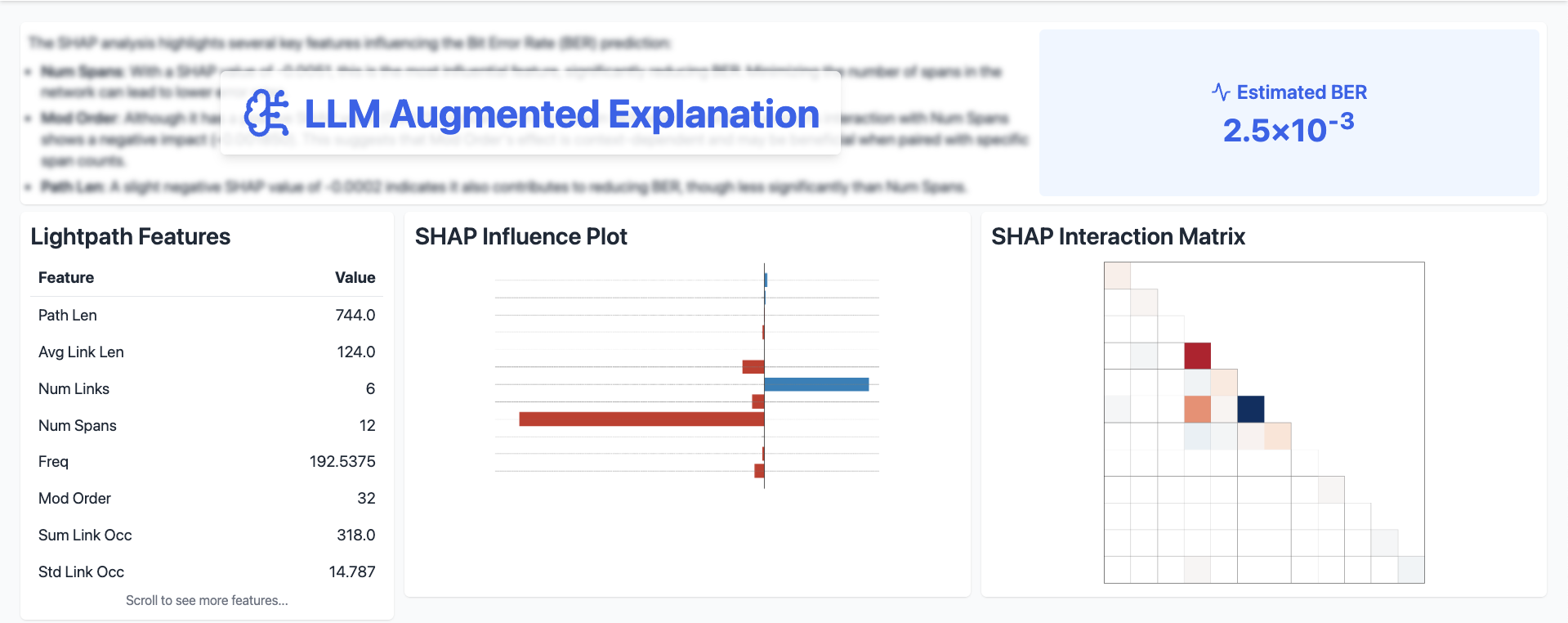} 
    \caption{Dashboard example: QoT estimation (top right) and LLM augmented explanation (top left) based on mutual feature interactions and influences (bottom part).}
    \label{fig:dashboard}
\end{figure}

\noindent
\textit{\uline{Dashboard:}} 
Experts need to inspect the output of AI/ML models before these outputs, or decisions based on them, are deployed in the network.
Fig. \ref{fig:dashboard} illustrates a dashboard example where information about features, SHAP influence, interaction matrix, and AI/ML model output values is augmented with the textual interpretations provided by our framework.

\section{Experimental and Empirical Results}
\label{sec: experimental and emprical results}

To evaluate the performance of the proposed framework, we began by training an \ac{XGB} regressor model using the dataset available in \cite{bergk2021ml} from which we selected a subset of 12 features that are the most indicative of signal quality and topological structure. 
The subset includes metrics related to path and link lengths, modulation formats, lightpath and connection line rates, source and destination node degrees, number of spans, and detailed link occupation statistics.
The dataset was split into training and test sets using a 90:10 ratio. 
The model's performance was evaluated using the test set, yielding a mean squared error (MSE) of 0.0000 and a mean absolute error (MAE) of 0.0001, both of which indicate high accuracy.
We randomly extracted 40 local SHAP explanations for our experiments\footnote{The GitHub repository\cite{llm4xai-repo} contains the prompt templates, the dataset including SHAP values and SHAP interaction values, as well as the corresponding \ac{LLM} explanations.}.

For the explanation augmentation module, we utilized a 4-bit quantized version of DeepSeek-R1 \ac{LLM}, consisting of 32.8 billion parameters, following the work of \cite{guo2025deepseek}.
We selected DeepSeek-R1 for its strong reasoning capabilities, which are essential for generating high-quality explanations.
To enable efficient local inference with relatively fast response times, we adopted quantization instead of distillation, a choice supported by findings in \cite{zhang2025reasoningmeetscompressionbenchmarking}, which show that quantization more effectively preserves reasoning performance in \acp{LLM}.

As a baseline, we employed a straightforward prompt asking the \ac{LLM} to explain the rationale behind the predictions made by the \ac{XGB} model based only on the SHAP feature influence values \cite{10742571, Ayoub_2025_icton}.
An example comparing the prompt designs and generated explanations for the baseline approach and proposed frameworks is shown in Table~\ref{tab:qualitative_comparison} in the Appendix. 
The baseline and the proposed prompting approaches require fewer than 1,500 input tokens. 
The \ac{LLM} output was also below 1,500 tokens, well below any token limit of modern \acp{LLM}.

In the final evaluation step, two human experts analyzed and assessed the generated interpretations. 
To ensure objectivity, the interpretations were presented in a randomized order, preventing the experts from identifying the prompting strategy that had generated each one.
Additionally, the experts conducted their evaluations independently, without access to each other’s assessments, thereby minimizing potential bias. Both experts are senior researchers with extensive experience in applying \ac{XAI} to optical network automation.
We adopt the human-centered evaluation metrics proposed in \cite{Ayoub_2025_icton}, which assess the LLM-generated interpretations across three dimensions:
(i) \emph{correctness}, measuring how faithfully the interpretation reflects the underlying explanation;
(ii) \emph{scope}, evaluating whether the interpretation, when correct, highlights the most critical aspects of the explanation\footnote{An interpretation may be accurate but still fail to emphasize the most relevant features.}; and
(iii)~\emph{usefulness}, gauging, when correct, how helpful the interpretation is in supporting human understanding.
Correctness and scope are rated using binary labels (e.g., correct/incorrect, in/out of scope). 
Usefulness is rated on a scale of 0 to 5, reflecting the extent to which the explanation helps the evaluator understand the model's behavior.
Table \ref{tab:prompt_comparison} reports the empirical results. Results for \emph{scope} and \emph{usefulness} are only considered when the expert evaluates the interpretation as correct. 
We also report the agreement among experts across \emph{correctness} and \emph{scope}, as well as the standard deviation of \emph{usefulness}.

Results indicate that the baseline strategy achieves an average correctness of 96.3\%, while the proposed framework yields a slightly higher correctness of 97.5\%.
Evaluator agreement on correctness is high for both strategies, at 93\% and 95\%, respectively.
These results confirm that the integration of a reasoning \ac{LLM}, mutual feature interaction data, and a structured prompt yields a significant performance improvement over methods available in the literature \cite{Ayoub_2025_icton}.
This is particularly evident in the scope of the explanations, where the proposed framework exhibits a notably higher score (95.0\%) compared to the baseline (88.8\%). 
This substantial improvement suggests that the use of mutual feature interactions helps the \ac{LLM} generate more comprehensive and contextually relevant explanations.

\begin{table}[t]
\caption{Results Showing the Average (Avg), Agreement (Agr), and Standard Deviation (Std).}
\label{tab:prompt_comparison}
\begin{tabularx}{\columnwidth}{X|c|c|c|c} 
\hline
\textbf{Metric} & \multicolumn{2}{c|}{\textbf{Baseline}} & \multicolumn{2}{c}{\textbf{Proposed}} \\
\cline{2-5}
 & \textbf{\textit{Avg}} & \textbf{\textit{Agr/Std}} & \textbf{\textit{Avg}} & \textbf{\textit{Agr/Std}} \\
\hline
Correctness & $96.3\%$ & $93\%$ / $-$ & $97.5\%$ & $95\%$ / $-$ \\
\hline
Scope       & $88.8\%$ & $83\%$ / $-$ & $95.0\%$ & $90\%$ / $-$ \\
\hline
Usefulness  & $3.77$   & $-$ / $1.5$  & $4.38$   & $-$ / $1.4$ \\
\hline
\multicolumn{5}{l}{\footnotesize $^{\mathrm{}}$ Avg = Average, Agr = Agreement, Std = Standard Deviation.}
\end{tabularx}
\end{table}

Regarding usefulness, the evaluations of experts further support the benefits of the proposed framework.
The baseline approach received an average usefulness rating of 3.77 (std 1.4), while the proposed framework achieved a higher rating of 4.38 (std 1.5).
This further supports the finding that the structured design and the explicit inclusion of SHAP mutual feature interaction values significantly improve the clarity and interpretability of the explanations.

\section{Conclusion}
\label{sec: conclusion}

This paper presents a framework designed to bridge the gap between complex explanations from \ac{XAI} methods and human-readable interpretations.
It leverages a moderately-sized \ac{LLM} with advanced reasoning capabilities, and a structured prompt that incorporates both individual SHAP feature influence values and mutual feature interaction data. 
We evaluated our proposed framework against a state-of-the-art prompting approach that utilizes SHAP influence values.
Both used a reasoning \ac{LLM} and were found to be effective, with the proposed framework consistently outperforming the basic approach.
These results highlight the importance of considering structured prompts with feature-interaction-aware context. While SHAP offers detailed insights, its cost in complex \ac{AI/ML} models invites exploring lighter methods like LIME or gradient-based approaches within this framework \cite{vale2022explainable}.
Ultimately, the adoption of this approach in operational settings can enable scalable interpretability of \ac{AI/ML} model outputs, improve transparency when using \ac{AI/ML} models in real-world deployments, and increase the overall trustworthiness of decisions made based on \ac{AI/ML} outputs.

\appendix
\setcounter{table}{0}
\renewcommand{\thetable}{A-\Roman{table}}

This appendix contains supplementary tables referenced in the main text. Table \ref{tab:prompt_structure} specifies the complete, multi-part prompt structure designed for our proposed framework, and Table \ref{tab:qualitative_comparison} shows a qualitative comparison of the generated explanations to highlight the practical differences in output quality between our approach and the state-of-the-art baseline.
\begin{table*}[htbp]
\centering
\scriptsize
\caption{Prompt Structure Design for Explanation Augmentation Module}
\label{tab:prompt_structure}
\renewcommand{\arraystretch}{0.5} 
\setlength{\tabcolsep}{2pt} 
\begin{tabularx}{\textwidth}{@{}l >{\RaggedRight\arraybackslash}X >{\RaggedRight\arraybackslash}X@{}}
\toprule
\textbf{Component} & \textbf{Description} & \textbf{Example Snippet from Prompt} \\ \midrule
\textbf{[Context]} &
  Defines the LLM's persona and the problem domain. \newline
  - \textbf{Role}: Interpreter of a SHAP explanation. \newline
  - \textbf{Target}: Bit Error Rate (BER) prediction for optical lightpaths. \newline
  - \textbf{Rules}: Specifies how to interpret positive/negative SHAP values. &
  \texttt{You are an interpreter of a SHAP explanation... predicting Bit Error Rate (BER)... Positive SHAP → Increases BER (undesired)} \\ \midrule
\textbf{[Task]} &
  Outlines the primary instructions for the LLM. \newline
  1. Identify the 2–3 most impactful features from SHAP values. \newline
  2. Describe each feature's value and its influence on BER. \newline
  3. Summarize findings with actionable insights to reduce BER. &
  \texttt{1. Identify the top 2 or 3 features... \newline 3. Summarize with a concise explanation covering... Practical insights to reduce BER} \\ \midrule
\textbf{[Interaction Handling]} &
  Provides conditional logic for complex or unexpected results. \newline
  - \textbf{Trigger}: If a feature's effect contradicts domain knowledge. \newline
  - \textbf{Action}: Analyze SHAP interaction values. to find the root cause. \newline
  - \textbf{Output}: Revise the conclusion to reflect interaction-driven effects. &
  \texttt{If a feature seems to reduce BER in a way that contradicts domain expectations: 1. Refer to the SHAP interaction values...} \\ \midrule
\textbf{[Input Explanation]} &
  The raw data provided to the LLM for a single instance. \newline
  - Includes the model's prediction (\texttt{ML Model Prediction}). \newline
  - Includes individual feature data (\texttt{Feature-wise breakdown}). \newline
  - Includes pairwise interaction data (\texttt{Feature Interaction Breakdown}). &
  \texttt{ Input Explanation: \newline ML Model Prediction: 0.00024... \newline Feature-wise breakdown: \newline - Num Spans: Min Value = 2.0, ...,Value= 13, SHAP Value= 0.0059 ... \newline Feature Interaction Breakdown: \newline - Num Spans \& Mod Order: SHAP Interaction Value = -0.000862 ...} \\ \midrule
\textbf{[Response Structure]} &
  Defines the required format for the LLM's output. \newline
  1. \textbf{Interpretation}: A detailed breakdown of key features and their effects. \newline
  2. \textbf{Conclusion}: A concise summary with actionable insights. &
  \texttt{Response Structure\newline 1. Interpretation ...\newline 2. Conclusion ...} \\ \bottomrule
\end{tabularx}
\end{table*}
\begin{table*}[htbp]
\centering
\scriptsize
\caption{Qualitative Comparison of Baseline and Proposed Frameworks for an Example Instance}

\label{tab:qualitative_comparison}
\renewcommand{\arraystretch}{0.5} 
\setlength{\tabcolsep}{2pt} 
\begin{tabularx}{1\textwidth}{@{}l >{\RaggedRight\arraybackslash}X >{\RaggedRight\arraybackslash}X@{}}
\toprule
\textbf{Component} & \textbf{Baseline Approach} & \textbf{Proposed Framework} \\ \midrule
\textbf{Prompt} &
  \texttt{Below is the output of a SHAP explanation from a machine learning model that predicts Bit Error Rate (BER) in optical networks. Please write a short explanation of what this means and which features seem most important... Feature-wise breakdown:
- Path Len: Min Value = 84.0, ..., SHAP Value = 0.0001
...} &
  \texttt{\#\# [Context]\newline You are an interpreter of a SHAP explanation...\newline [Task]\newline 1. Identify the top 2–3 features...\newline \#\# [Interaction Handling]...} \newline (see Table~\ref{tab:prompt_structure} for full details). \\
\midrule
\textbf{Generated Explanation} &
  \texttt{\textbf{Interpretation}
  \newline The SHAP (SHapley Additive exPlanations) output explains how different features contribute to the model's prediction of Bit Error Rate (BER). In this case, the model predicts a BER of approximately 0.00045. \newline\mbox{}\newline
  \textbf{Key Insights:} \newline
  1. \textbf{Most Important Features}: \newline
  \quad - \textbf{Num Spans}: This feature has the largest negative contribution (-0.0033), meaning that increasing the number of spans in the network reduces the predicted BER. \newline
  \quad - \textbf{Mod Order (Modulation Order)}: This feature has the largest positive contribution (+0.0028), indicating that higher modulation orders increase the predicted BER. \newline
  2. \textbf{Other Notable Features}: \newline
  \quad - \textbf{Path Len}: Slightly increases the predicted BER (+0.0001). \newline
  \quad - \textbf{Freq, Sum Link Occ, and others}: These features have negligible contributions to the prediction, as their SHAP values are close to zero. \newline\mbox{}\newline
  \textbf{Summary:} \newline
  The model's prediction of a low BER (0.00045) is primarily influenced by the number of spans in the network and the modulation order used. More spans tend to reduce errors, while higher modulation orders increase them.} &
  \texttt{\textbf{Interpretation} \newline
  The SHAP analysis reveals that the Bit Error Rate (BER) prediction of approximately 0.00045 is significantly influenced by two key features: \newline
  1. \textbf{Modulation Order (Mod Order)}: With a SHAP value of +0.0028, Mod Order has the most substantial positive impact on increasing BER. The current value is set to its maximum at 64, which contributes to higher error rates. \newline
  2. \textbf{Number of Spans}: This feature has the largest negative SHAP value (-0.0033), indicating that fewer spans reduce BER. The current span count is 6.0. 
  Additionally, a notable interaction exists between Mod Order and Number of Spans (SHAP interaction value -0.001591). This suggests that when both factors are present, their combined effect on reducing BER is more pronounced than the sum of their individual effects. \newline\mbox{}\newline
  \textbf{Conclusion} \newline
  To optimize BER reduction: \newline
  1. \textbf{Reduce Modulation Order}: Lowering Mod Order from 64 to a feasible minimum (e.g., closer to 16) could significantly decrease BER. \newline
  2. \textbf{Optimize Number of Spans}: Further reducing spans below 6, if technically viable, would enhance BER improvement, especially when combined with lower Mod Order. \newline\mbox{}\newline
  By addressing these factors, particularly focusing on lowering Mod Order and maintaining fewer spans, the system can achieve a more substantial reduction in BER.
  } 
  \\
  \bottomrule
\end{tabularx}
\end{table*}

\clearpage
\acrodef{AI}{artificial intelligence}
\acrodef{AI/ML}{artificial intelligence and machine learning}
\acrodef{ANN}{artificial neural network}
\acrodef{ASE}{amplified spontaneous emissions}
\acrodef{CUT}{channel under test}
\acrodef{DRL}{deep reinforcement learning}
\acrodef{EGN}{enhanced Gaussian noise}
\acrodef{EON}{elastic optical network}
\acrodef{GT}{ground truth}
\acrodef{GSNR}{generalized signal-to-noise ratio}
\acrodef{KDE}{kernel density estimation}
\acrodef{LLM}{large language model}
\acrodef{LP}{lightpath}
\acrodef{MAE}{mean absolute error}
\acrodef{MF}{modulation format}
\acrodef{ML}{machine learning}
\acrodef{NLI}{non-linear impairments}
\acrodef{QoT}{quality of transmission}
\acrodef{RMSA}{routing, modulation and spectrum assignment}
\acrodef{XGB}{XGBoost}
\acrodef{IBN}{intent-based networking}
\acrodef{BER}{bit error rate}
\acrodef{XAI}{explainable artificial intelligence}
\acrodef{ADON}{autonomous driving optical network}
\acrodef{SHAP}{SHapley Additive exPlanations}
\IEEEtriggeratref{16}
\bibliographystyle{IEEEtran}
\bibliography{references}

@article{Alhammadi2024,
author = {Alhammadi, Abdulraqeb and Shayea, Ibraheem and El-Saleh, Ayman A. and Azmi, Marwan Hadri and Ismail, Zool Hilmi and Kouhalvandi, Lida and Saad, Sawan Ali},
title = {Artificial Intelligence in {6G} Wireless Networks: Opportunities, Applications, and Challenges},
journal = {International Journal of Intelligent Systems},
volume = {2024},
number = {1},
pages = {8845070},
doi = {https://doi.org/10.1155/2024/8845070},
year = {2024}
}

@inproceedings{ayoub2024xrl,
  title={Towards explainable reinforcement learning in optical networks: The {RMSA} use case},
  author={Ayoub, Omran and Natalino, Carlos and Monti, Paolo},
  booktitle={Optical Fiber Communications Conference and Exhibition (OFC)},
  pages={W4I.6},
  year={2024},
}

@article{Etezadi:23,
author = {Ehsan Etezadi and Carlos Natalino and Renzo Diaz and Anders Lindgren and Stefan Melin and Lena Wosinska and Paolo Monti and Marija Furdek},
journal = {J. Opt. Commun. Netw.},
number = {10},
pages = {E86--E96},
publisher = {Optica Publishing Group},
title = {Deep reinforcement learning for proactive spectrum defragmentation in elastic optical networks},
volume = {15},
month = {Oct},
year = {2023},
doi = {10.1364/JOCN.489577}
}

@article{Natalino:24,
author = {Carlos Natalino and Ashkan Panahi and Nasser Mohammadiha and Paolo Monti},
journal = {J. Opt. Commun. Netw.},
number = {2},
pages = {A169--A179},
publisher = {Optica Publishing Group},
title = {{AI/ML-as-a-Service} for optical network automation: use cases and challenges {{Invited}}},
volume = {16},
month = {Feb},
year = {2024},
doi = {10.1364/JOCN.500706}
}

@inproceedings{fawaz,
  title={Reducing Complexity and Enhancing Predictive Power of {ML}-Based Lightpath {QoT} Estimation via {SHAP}-Assisted Feature Selection},
  author={Fawaz, Hussein and Arpanaei, Farhad and Andreoletti, Davide and Sbeity, Ihab and Hern{\'a}ndez, Jos{\'e} Alberto and Larrabeiti, David and Ayoub, Omran},
  booktitle={International Conference on Optical Network Design and Modeling (ONDM)},
  year={2024},
}

@inproceedings{ayoub2022quantifying,
  title={Quantifying features’ contribution for {ML}-based quality-of-transmission estimation using explainable {AI}},
  author={Ayoub, Omran and Andreoletti, Davide and Troia, Sebastian and Giordano, Silvia and Bianco, Andrea and Rottondi, Cristina},
  booktitle={European Conference on Optical Communication (ECOC)},
  pages={We3B.4},
  year={2022},
}

@article{allogba2022machine,
  title={Machine-learning-based lightpath {QoT} estimation and forecasting},
  author={Allogba, St{\'e}phanie and Aladin, Sandra and Tremblay, Christine},
  journal={Journal of Lightwave Technology},
  volume={40},
  number={10},
  pages={3115--3127},
  year={2022}
}

@article{rottondi2018machine,
  title={Machine-learning method for quality of transmission prediction of unestablished lightpaths},
  author={Rottondi, Cristina and Barletta, Luca and Giusti, Alessandro and Tornatore, Massimo},
  journal={Journal of Optical Communications and Networking},
  volume={10},
  number={2},
  pages={A286--A297},
  year={2018},
  publisher={OSA}
}

@article{El-Hajj2025,
author = {El-Hajj, Mohammed},
title = {Enhancing Communication Networks in the New Era with Artificial Intelligence: Techniques, Applications, and Future Directions},
journal = {Network},
volume = {5},
year = {2025},
number = {1},
issn = {2673-8732},
doi = {10.3390/network5010001}
}

@Article{nazi2024largelanguagemodelshealthcare,
AUTHOR = {Nazi, Zabir Al and Peng, Wei},
TITLE = {Large Language Models in Healthcare and Medical Domain: A Review},
JOURNAL = {Informatics},
VOLUME = {11},
YEAR = {2024},
NUMBER = {3},
ARTICLE-NUMBER = {57},
ISSN = {2227-9709},
DOI = {10.3390/informatics11030057}
}

@article{ahmed2024explainable,
  title={Explainable {AI}-based innovative hybrid ensemble model for intrusion detection},
  author={Ahmed, Usman and Jiangbin, Zheng and Almogren, Ahmad and others},
  journal={Journal of Cloud Computing},
  volume={13},
  number={1},
  pages={150},
  year={2024},
  publisher={Springer},
  doi={https://doi.org/10.1186/s13677-024-00712-x}
}

@article{bikkasani2024ai,
  title={{AI}-Driven {5G} Network Optimization: A Comprehensive Review of Resource Allocation, Traffic Management, and Dynamic Network Slicing},
  author={Bikkasani, Dheeraj Chandra and Yerabolu, Mohan Reddy},
  journal={American Journal of Artificial Intelligence},
  volume={8},
  number={2},
  pages={55--62},
  year={2024},
  doi={10.11648/j.ajai.20240802.14}
}

@misc{nyswa2025future,
  author       = {{New York State Wireless Association}},
  title        = {The Future of Telecom Infrastructure: The Network on Which {AI} Will Be Built},
  howpublished = {\url{https://nyswa.org/the-future-of-telecom-infrastructure-the-network-on-which-ai-will-be-built/}},
  note         = {Accessed: 2025-04-30},
  year         = {2025},
  month        = mar
}

@ARTICLE{Musumeci_2019_survey,
  author={Musumeci, Francesco and Rottondi, Cristina and Nag, Avishek and Macaluso, Irene and Zibar, Darko and Ruffini, Marco and Tornatore, Massimo},
  journal={IEEE Communications Surveys \& Tutorials}, 
  title={An Overview on Application of Machine Learning Techniques in Optical Networks}, 
  year={2019},
  volume={21},
  number={2},
  pages={1383-1408},
  doi={10.1109/COMST.2018.2880039}}

@article{wu2022knowledge,
  title={Knowledge-powered explainable artificial intelligence for network automation toward {6G}},
  author={Wu, Yulei and Lin, Guozhi and Ge, Jingguo},
  journal={IEEE network},
  volume={36},
  number={3},
  pages={16--23},
  year={2022},
  publisher={IEEE},
  doi={10.1109/MNET.005.2100541}
}

@article{wang2021applications,
  title={Applications of explainable {AI} for {6G}: Technical aspects, use cases, and research challenges},
  author={Wang, Shen and Qureshi, M Atif and Miralles-Pechuan, Luis and Huynh-The, Thien and Gadekallu, Thippa Reddy and Liyanage, Madhusanka},
  journal={arXiv preprint arXiv:2112.04698},
  year={2021},
  doi={10.48550/arXiv.2112.04698}
}

@article{ayoub2022towards,
  title={Towards explainable artificial intelligence in optical networks: the use case of lightpath {QoT} estimation},
  author={Ayoub, Omran and Troia, Sebastian and Andreoletti, Davide and Bianco, Andrea and Tornatore, Massimo and Giordano, Silvia and Rottondi, Cristina},
  journal={Journal of Optical Communications and Networking},
  volume={15},
  number={1},
  pages={A26--A38},
  year={2022},
  publisher={IEEE},
  doi={10.1364/JOCN.470812}}

@article{dutta2021challenge,
  title={The challenge of zero touch and explainable {AI}},
  author={Dutta, Biswadeb and Krichel, Andreas and Odini, Marie-Paule},
  journal={Journal of ICT Standardization},
  volume={9},
  number={2},
  pages={147--158},
  year={2021},
  publisher={River Publishers},
  doi={10.13052/jicts2245-800X.925}
}

@article{vale2022explainable,
  title={Explainable artificial intelligence ({XAI}) post-hoc explainability methods: Risks and limitations in non-discrimination law},
  author={Vale, Daniel and El-Sharif, Ali and Ali, Muhammed},
  journal={AI and Ethics},
  volume={2},
  number={4},
  pages={815--826},
  year={2022},
  publisher={Springer},
  doi={10.1007/s43681-022-00142-y}
}

@article{weber2023beyond,
  title={Beyond explaining: Opportunities and challenges of {XAI}-based model improvement},
  author={Weber, Leander and Lapuschkin, Sebastian and Binder, Alexander and Samek, Wojciech},
  journal={Information Fusion},
  volume={92},
  pages={154--176},
  year={2023},
  publisher={Elsevier},
  doi={10.1016/j.inffus.2022.11.013}
}

@inproceedings{hudon2021explainable,
  title={Explainable artificial intelligence ({XAI}): how the visualization of {AI} predictions affects user cognitive load and confidence},
  author={Hudon, Antoine and Demazure, Th{\'e}ophile and Karran, Alexander and L{\'e}ger, Pierre-Majorique and S{\'e}n{\'e}cal, Sylvain},
  booktitle={Information Systems and Neuroscience: NeuroIS Retreat},
  pages={237--246},
  year={2021},
  organization={Springer},
  doi={10.1007/978-3-030-88900-5_27}
}

@inproceedings{lundberg2017unifiedapproachinterpretingmodel,
 author = {Lundberg, Scott M and Lee, Su-In},
 booktitle = {Advances in Neural Information Processing Systems},
 editor = {I. Guyon and U. Von Luxburg and S. Bengio and H. Wallach and R. Fergus and S. Vishwanathan and R. Garnett},
 pages = {},
 title = {A Unified Approach to Interpreting Model Predictions},
 volume = {30},
 year = {2017}
}

@INPROCEEDINGS{adanza2024intent,
  author={Adanza, Daniel and Natalino, Carlos and Gifre, Lluis and Muñoz, Raul and Alemany, Pol and Monti, Paolo and Vilalta, Ricard},
  booktitle={IEEE 10th International Conference on Network Softwarization (NetSoft)}, 
  title={{IntentLLM}: An {AI} Chatbot to Create, Find, and Explain Slice Intents in {TeraFlowSDN}}, 
  year={2024},
  volume={},
  number={},
  pages={307-309},
  doi={10.1109/NetSoft60951.2024.10588917}}

@article{bergk2021ml,
  title={{ML}-assisted {QoT} estimation: a dataset collection and data visualization for dataset quality evaluation},
  author={Bergk, Geronimo and Shariati, Behnam and Safari, Pooyan and Fischer, Johannes K},
  journal={Journal of Optical Communications and Networking},
  volume={14},
  number={3},
  pages={43--55},
  year={2021},
  publisher={Optical Society of America},
  doi={10.1364/JOCN.442733}
}

@article{guo2025deepseek,
  title={Deepseek-r1: Incentivizing reasoning capability in {LLMs} via reinforcement learning},
  author={Guo, Daya and Yang, Dejian and Zhang, Haowei and Song, Junxiao and Zhang, Ruoyu and Xu, Runxin and Zhu, Qihao and Ma, Shirong and Wang, Peiyi and Bi, Xiao and others},
  journal={arXiv preprint arXiv:2501.12948},
  year={2025},
  doi={10.48550/arXiv.2501.12948}
}

@misc{zeng2024enhancinginterpretabilityshapvalues,
      title={Enhancing the Interpretability of {SHAP} Values Using Large Language Models}, 
      author={Xianlong Zeng},
      year={2024},
      eprint={2409.00079},
      archivePrefix={arXiv},
      primaryClass={cs.HC},
      url={https://arxiv.org/abs/2409.00079}, 
}

@misc{zhang2025reasoningmeetscompressionbenchmarking,
      author={Nan Zhang and Yusen Zhang and Prasenjit Mitra and Rui Zhang},
      year={2025},
title={When Reasoning Meets Compression: Benchmarking Compressed Large Reasoning Models on Complex Reasoning Tasks},
      eprint={2504.02010},
      archivePrefix={arXiv},
      primaryClass={cs.LG},
      url={https://arxiv.org/abs/2504.02010}, 
}

@ARTICLE{10742571,
  author={Mekrache, Abdelkader and Mekki, Mohamed and Ksentini, Adlen and Brik, Bouziane and Verikoukis, Christos},
  journal={IEEE Communications Magazine}, 
  title={On Combining {XAI} and {LLMs} for Trustworthy Zero-Touch Network and Service Management in {6G}}, 
  year={2025},
  volume={63},
  number={4},
  pages={154-160},
  doi={10.1109/MCOM.002.2400276}}

@article{Ayoub_2025_icton,
  title     = {Natural Language Interpretability for {ML}-Based {QoT} Estimation via Large Language Models},
  author    = {Ayoub, Omran and Troia, Sebastian and Natalino, Carlos and Rottondi, Cristina and Andreoletti, Davide and Lelli, Francesco and Giordano, Silvia and Monti, Paolo},
  journal   = {International Conference on Transparent Optical Networks (ICTON)},
  pages     = {Tu.C2.4},
  year      = {2025},
}

@misc{llm4xai-repo,
  author       = {Kiarash Rezaei},
  title        = {{LLM}-Augmented {XAI} with Mutual Feature Interactions: The {QoT} Estimation Use Case},
  howpublished = {\url{https://github.com/kiarashRezaei/llm-for-xai-qotEstimation}}
}

\end{document}